\documentstyle[prl,aps]{revtex}

\begin{document}


\title{How Much Information can be Obtained by a Quantum Measurement ?}
\author{S. Massar$^a$ and S. Popescu$^{bc}$}
\address{$^a$ Service de Physique Th\'eorique, 
Universit\'e Libre de Bruxelles,\\
CP 225, Bvd. du Triomphe,
B1050 Bruxelles, Belgium.\\
$^b$ Isaac Newton Institute, University of Cambridge, Cambridge CB3
0EH, U.K. \\
$^c$ BRIMS, Hewlett-Packard Labs., Bristol BS12 SQZ, U.K.
}

\maketitle

\begin{abstract}
How much information about an unknown quantum state can be obtained by a
measurement? We propose a model independent answer: the information
obtained is equal to the minimum entropy of the outputs of the
measurement, where the minimum is taken over all measurements which
measure the same ``property'' of the state. 
This minimization is necessary because
the measurement outcomes can be redundant, and this redundancy must be
eliminated.
We show that this minimum entropy is
less or equal than the von Neumann entropy of the 
unknown states.
That is a measurement can extract at most one meaningful bit
from every qubit 
carried by the unknown states.
\end{abstract}



\newsavebox{\channel}
\savebox{\channel}{\begin{picture}(0,0)
\put(0,-40){\framebox(80,80){}}
\thicklines
\put(80,0){\vector(1,0){120}}
\thinlines
\put(200,-40){\framebox(80,80){}}
\put(280,0){\vector(2,1){60}}
\put(280,0){\vector(4,1){60}}
\put(280,0){\vector(1,0){60}}
\put(280,0){\vector(4,-1){60}}
\put(280,0){\vector(2,-1){60}}
\end{picture}}

\newsavebox{\channeltwo}
\savebox{\channeltwo}{\begin{picture}(0,0)
\put(0,-40){\framebox(80,80){}}
\thicklines
\put(80,0){\vector(1,0){80}}
\thinlines
\put(160,-40){\framebox(80,80){}}
\put(240,0){\vector(2,1){60}}
\put(240,0){\vector(4,1){60}}
\put(240,0){\vector(1,0){60}}
\put(240,0){\vector(4,-1){60}}
\put(240,0){\vector(2,-1){60}}
\end{picture}}

\newsavebox{\annchannel}
\savebox{\annchannel}
{\begin{picture}(0,0)
\put(0,0){\usebox{\channel}}
\put(20,0){\makebox(40,20)[t]{\bf Quantum }} 
\put(20,-30){\makebox(40,20)[t]{\bf Source }} 
\put(220,0){\makebox(40,20)[t]{\bf Measuring }} 
\put(220,-30){\makebox(40,20)[t]{\bf Device }} 
\put(90,-90){\makebox(100,80)[t]{
\begin{minipage}{100pt}\begin{center}
states $\psi_i$ \\
probabilities $p_i$\\
$\rho\! =\! \sum_i p_i |\psi_i><\psi_i|$\\
$I_{input}\! =\! - Tr \rho \ln \rho$
\end{center}
\end{minipage}
}}
\put(350,-50){\makebox(100,80)[t]{
\begin{minipage}{110pt}
N outcomes \\
probabilities $p_j$\\
$I_{output}\! =\! - \sum_j p_j \ln p_j$
\end{minipage}
}}
\end{picture}}

\newsavebox{\classcomm}
\savebox{\classcomm}{\begin{picture}(0,0)
\put(0,-40){\framebox(80,80){}}
\put(20,0){\makebox(40,20)[t]{\bf Classical }} 
\put(20,-30){\makebox(40,20)[t]{\bf Source }} 
\put(80,0){\vector(1,0){40}}
\put(120,0){\usebox{\channeltwo}}
\put(140,0){\makebox(40,20)[t]{\bf Quantum }} 
\put(140,-30){\makebox(40,20)[t]{\bf Source }} 
 \put(300,0){\makebox(40,20)[t]{\bf Measuring }} 
\put(300,-30){\makebox(40,20)[t]{\bf Device }} 
\put(190,-90){\makebox(100,80)[t]{
\begin{minipage}{100pt}\begin{center}
states $\psi_i$ 
\end{center}
\end{minipage}
}}
\put(400,-120){\makebox(100,80)[t]{
\begin{minipage}{110pt}
output
\end{minipage}
}}
\end{picture}
}


\section{Introduction}

Quantum mechanics has at its core a fundamental statistical aspect. 
Suppose you 
are given a single quantum particle in a state $|\Psi\rangle$ unknown 
to you. There 
is no way to find what $|\Psi\rangle$ 
is - to find it out you need an infinite 
ensemble of quantum particles, all prepared in the same state. Indeed, the 
different properties which characterize the state are, in general,
complementary to 
one another; measuring one disturbes the rest. Only if an infinite 
ensemble is 
given can one find out the state.  But 
infinite ensembles don't exist in practice. Given a finite ensemble of 
identically prepared particles, how 
well can one estimate the state? The problem is a fundamental one for 
understanding the very basis of quantum mechanics. It has been investigated by 
many authors, see for instance \cite{Hel}\cite{H2}, 
and it constitutes probably the oldest problem in what is at 
present called ``quantum information". Here we approach this 
problem from a new 
point of view which, we think, leads to a deeper understanting.

What is the optimal way to estimate the quantum state given a finite 
ensemble? 
As such the question is not well posed. Indeed, since we cannot completely 
determine the state , i.e. completely determine all its properties, we 
must decide which particular property we want to determine.
For an ensemble of spins, for example, 
estimating as well as possible the mean value of the $z$ spin component is, 
obviously, a different question than estimating as well as possible the mean 
value of the $x$ spin component. 

But things are in fact even more complicated. The apparent benign words ``as 
well as possible" in the previous paragraph are not well defined. Indeed, ``as 
well as possible" actually means ``as well as possible given a 
specific measure 
of what ``well" means". Obviously, one can imagine many different 
measures. For 
example, suppose that a source emits  states $|\psi_i\rangle$ with 
 probability 
$p_i$. The problem is to design a measurement at the 
end of which we must guess which state was emitted.
Let the guess be $|\phi_j^{guess}\rangle $,
and let the measure of success (fidelity) be 
\begin{equation}
F_{ij}=|\langle \phi_j^{guess}|\psi_i\rangle |^2\ ,\label{1}
\end{equation}
i.e. the absolute value square of the scalar product 
in between the true state 
$|\psi_i\rangle$ and the guess $|\phi_j^{guess}\rangle$. 
The goal is to optimize 
the measurement such that it yields the highest average fidelity
\begin{equation}
F=\sum_{i,j} p_i F_{ij}p(j|i)\ .\label{2}
\end{equation}
where $p(j|i)$ is the probability to make guess $j$ if the state is 
$|\psi_i\rangle $.
On the other hand, one can imagine another fidelity function, such as
\begin{equation}F'_{ij}=|\langle \phi_j^{guess}|\psi_i\rangle |^4\ .
\label{3}\end{equation}
Or one could try to optimize the mutual information
\begin{equation}I=
- \sum_i p_i \ln p_i + \sum_j p_j \sum_i p(i|j)\ln  p(i|j)
\label{4}\end{equation} 
or any other measure. 

The important point to notice about the above different problems is 
that the 
different fidelities (2-4) not only define different scales according 
to which 
we measure the degree of success in estimating the state, 
but also, implicitly, 
define which property of the state we are  actually estimating.
If all the different fidelities  where to
lead to the same optimal 
measurements, we could say that we learn the same property 
about the state but 
just expressed in a different way. However the different fidelities
will in general lead to different optimal 
measurements which  means that in each case we learn a different property 
about the system.

To summarize, in general each particular estimation problem is  completely 
different from the other, they measure different properties 
and their degree of 
success is measured on different scales, with the scales also defining 
implicitly what exactly is the property we estimate.

That one can learn different properties  
is a fact of life inherent to quantum mechanics. But 
there is no reason not to use the same scale to gauge how
successful we have been in learning the property we decided  to measure.
The aim of this paper is to propose such a universal scale, and in the
process to introduce a novel approach to quantum state estimation.

\section{Main idea}

The central point of our approach starts from a simple but fundamental
question: what do we actually learn from a measurement on a state?
Let us illustrate this question by an example. We shall contrast two
situations. Consider a source which emits spin 1/2 particles. In the
first case the particles are polarized with equal probability
along either the $+z$ 
($|\uparrow_z\rangle $) or $-z$  ($|\downarrow_z\rangle $)
directions.
In the
second case the states are polarized along random directions uniformly
distributed on the sphere. Suppose we want to identify the states as
well as possible according to the fidelity eq. (\ref{2}). In the first
case it is obvious that a measurement along $\sigma_z$ perfectly
identifies the state, hence the fidelity is $F=1$. In the second case,
it has been shown \cite{MP} that the measurement along $\sigma_z$ is also
optimal. But in this case the states cannot be identified perfectly,
and the fidelity is only $F=2/3$. 

Nevertheless the two situations seem extremely similar. In both cases
we perform the same measurement. And in 
 both cases before we
perform the measurement we know that the outcomes of the measurement
are either $+1$ or $-1$, and the a priori probabilities of the two
outcomes are equal. When we perform the measurement this uncertainty
is resolved. Hence in both cases the measurement yields 1 bit of
information. Our main idea is to interpret this quantity as the
information we extract from the state. Incidentally we note that in
both cases this information (the Shannon information of the outcomes)
equals the von Newmann entropy of the unknown states (both are equal
to 1).

This idea might seem
paradoxical at first sight because in one case we completely recognize
the state whereas in the
other case we recognize it badly. To understand let us introduce a
classical source that decides which quantum state is emitted from the
quantum source (see figure 1). 
In the first case the classical source must only
specify one bit (either $+z$ or $-z$) to determine which state is
emitted. In the second case it must provide a direction
$\underline{n}_{in}$ (ie. an infinite number  of bits) 
in order to specify the state $|\uparrow_{\underline{n}_{in}}\rangle $.
In both cases one extracts one bit of information. In the first case
this means that the classical information supplied by the source is
completely recovered. In the second case  information is lost.
However it is now clear that the loss does not occur during the
measurement, but during the first
step, where classical information is converted into quantum.

\begin{figure}
\begin{picture}(300,200)
\put(10,100){\usebox{\classcomm}}
\end{picture}
\caption{Chain of events leading to a quantum state estimation
problem. The classical source specifies which state should be
sent. The quantum source then emits the corresponding state. Finally the
measuring device tries to identify the emitted state.}
\end{figure}
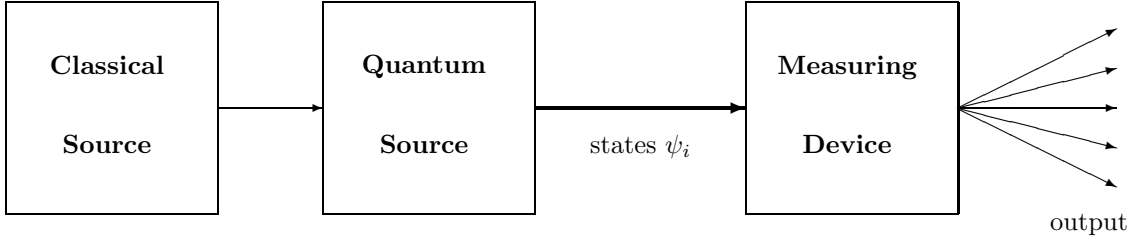

To summarize, 
the quantum state estimation problem as presented in figure 1 consists
of a chain of events which starts with a classical source that tells
the quantum source what state to emit, and ends with the measurement.
The fidelity measures the overall performance of the  chain
since it
is proportional to 
the scalar product
$\underline{n}_{in} . \underline{n}_{guess}$.
On the other hand the number of bits in the output characterizes how much
information is extracted by the measurement. Therefore in this article
we shall focus on the latter quantity.

\section{Main Result}

The preceding discussion suggests that the Shannon information of the
outcomes 
\begin{equation}
I_{output}^S = - \sum_j p_j \ln p_j\ ,
\end{equation}
where $p_j= \sum_i p(j|i)$ is the probability of outcome $j$, 
measures how much information is extracted from the state.
This idea however has to be refined.

The main problem is that there 
may be redundancies in the outputs of the measurement.  As 
a trivial example, a measurement could
be accompanied by the flip of a coin, and the outcomes of the
measurement would consist of both the outcomes of the measurement
proper and the outcomes of the coin flip. This adds one bit to the
entropy of the outputs without telling anything about the system. 
In less trivial examples involving POVM's and ancillas, redundancies
can arise in a less obvious way, and it is not immediate how they can
be identified and eliminated.

Our main result is that no matter what property of the system one
wants to measure, when the redundancy is eliminated, the remaining
Shannon information of the outputs has a universal upper bound which
is the von Neumann entropy of the quantum source:
\begin{equation} I_{output}^S (no\ redundancy) \leq I^{VN}_{input}\ ,
\label{S}\end{equation}
where $I^{VN}_{input} = -Tr \rho \ln \rho$ is the Shannon information of the
quantum source and $\rho$ is the density matrix of the quantum source
$\rho = \sum_i p_i |\psi_i\rangle \langle \psi_i|$.

One does not always attain equality in eq. (\ref{S}). Indeed some
questions are more informative about the system then others. Less
informative questions can be answered by measurements whose output
entropy is smaller. More informative questions require measurements
with more entropy. But the most detailed questions can always be
answered in $I^{VN}_{input}$ bits.

\section{Strategy}

The main problem we face in deriving eq. (\ref{S}) is to eliminate the
redundancy. In order to do this we shall proceed in several steps.
\begin{enumerate}
\item The first step is to decide which property we are interested
in. We may fix the property directly (for instance decide to measure
the average of $\sigma_z$) or implicitly by choosing a fidelity. In
the rest of the this paper we shall adopt the second approach.

\item We then look at optimal measurements, that is measurements
which maximize the fidelity. In general there is an entire class of
such measurements.

\item We perform a second optimization. Namely among the optimal
measurements  we look for
the measurements which minimize $I_{output}^S$. 
\end{enumerate}

This double optimization strategy has already been considered for some
particular cases in
\cite{DBE}\cite{LPT}\cite{LPT2}. 

One expects that this strategy yields measurements which have no
spurious redundancy. However as we will find out later through some
examples, redundancies cannot be completely eliminated by the above
procedure and we will have to further modify it.

These further modifications are motivated by 
the classical and quantum theory of
information\cite{Sh}\cite{S} which 
suggest the idea of
performing measurements on blocks of quantum states, rather than on
individual particles. Thus we shall allow the measuring device to
accumulate a large number $L$ of input states before making a
collective measurement on the $L$ states simultaneously. It is in the
context of these collective measurements that we make the two
optimizations (points 2 and 3 above) and thereby eliminate the
spurious redundancies. 

We want to emphasize that 
this procedure cannot increase the fidelity since
the subsequent particles are completely uncorrelated.
However by considering measurements on large blocks 
we can hope to reduce the redundancy of the measurement, ie. the
entropy of the outcomes, by making ``better use'' of each outcome.

Two technicalities have to be taken into account. First of all we 
must take care not to modify the definition of fidelity as we go from
measurements on single particles to block measurements.
That is the fidelity must still be the fidelity of each state
individually, rather than the fidelity for the whole block.
Second we should not require the measurement 
to absolutely maximize the fidelity, since then using block
measurements does not help to reduce the entropy (this follows once
more from the fact that the subsequent states are completely
uncorrelated). However, following the ideas of information theory, we
shall 
only require that the measurement has a fidelity approaching
arbitrarily closely the optimum.
In this framework we shall prove eq. (\ref{S}).

To summarize, there
is no best way of estimating an
unknown quantum state. Different measurements will learn about
different properties of the 
state, and it is up to us to choose which property we want to learn
about. However once we fix the property we want to learn about, we
show that quantitatively one cannot learn more than $I_{input}^{VN} = -
Tr \rho \ln \rho$ bits about this property. That is a measurement can
extract at most one meaningful bit from each qubit coming from the
source.

\section{Examples}\label{example}

Before embarking on a proof of our result, we give two examples which
illustrate the main points that must be taken into account in the proof.

In the first example there are two possible input states 
$|\psi_1\rangle  = \alpha
|\uparrow\rangle  + \beta |\downarrow\rangle $ and $|\psi_2\rangle  = \alpha
|\uparrow\rangle  - \beta |\downarrow\rangle $ which occur with equal
probability. The density matrix of the source is $\rho = \alpha^2 
|\uparrow\rangle \langle \uparrow| + \beta^2 |\downarrow\rangle 
\langle \downarrow|$ which is
different from the identity for $\alpha \neq \beta$ Therefore the von
Newmann entropy of the input states $I_{input}^{VN} < 1$ qubit.

In this example we use a fidelity defined as follows:
after each measurement one must guess whether
the state is $|\psi_1\rangle $ or  $|\psi_2\rangle $. In case of a 
correct guess one
receives a score of $+1$, and for an incorrect guess one receives a
score of $-1$. The aim is to maximize the average score.
The techniques of section \ref{fidelity} can be used to show that the optimal
measurement is a von Neumann measurement of $\sigma_x$, see figure 2. The two
outcomes of this measurement occur with equal probability, and hence
$I_{output}^S=1 > I_{input}^{VN}$.

\begin{figure}
\begin{picture}(110,60)
\thicklines
\put(60,20){\vector(3,2){42}}
\put(105,45){$\psi_1$}
\put(60,20){\vector(-3,2){42}}
\put(5,45){$\psi_2$}
\thinlines
\put(60,20){\vector(1,0){50}}
\put(105,25){$\uparrow_x$}
\put(60,20){\vector(-1,0){50}}
\put(5,25){$\downarrow_x$}
\end{picture}
\caption{The two input states $|\psi_1\rangle , |\psi_2\rangle =
\alpha|\uparrow\rangle \pm \beta | \downarrow \rangle$. The
optimal measurement is a measurement of the spin in the $x$ direction.}
\end{figure}
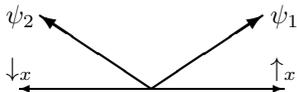

In this example, a natural first step in 
eliminating the redundancy is to project blocks of
input states onto their
probable subspace\cite{S}\cite{JS}. This projection succeeds with arbitrarily
high probability, and affects the input states arbitrarily
little. But it reduces the dimensionality of the Hilbert space of
the input states from $2^N$ to $2^{N I_{input}^{VN}}$. Hence if we can
prove that there is a von-Newman measurement restricted to the
probable subspace that is optimal, we will have proved our
claim. However the construction of such a von-Newmann measurement is
non trivial, as is illustrated in the next example.

In our second example 
 there is no ``most probable''
subspace because the density matrix of the inputs is completely
random. In this example  there are three input states
$|\psi_1\rangle  = |\uparrow\rangle $, $| \psi_2\rangle  ={1 \over 2} 
|\uparrow\rangle 
+{\sqrt{3} \over 2}|\downarrow\rangle $,  $\psi_3\rangle  ={1 \over 2}
|\uparrow\rangle 
-{\sqrt{3} \over 2}|\downarrow\rangle $, each occurring with equal probability
$p_i = 1/3$. The density matrix of these states is $\rho = I/2$ and
their entropy is $I_{input}^{VN} = 1$ qubit.
The fidelity is defined as above:
after the measurement one must guess which was the input state. If the
guess is correct one scores $+1$ point, if the guess is incorrect, one
scores $-1$ points. The aim is to maximize the average score (fidelity).

\begin{figure}[h]
\begin{picture}(110,110)
\thicklines
\put(50,50){\vector(0,1){50}}
\put(55,95){$\psi_1$}
\put(50,50){\vector(3,-2){42}}
\put(87,30){$\psi_2$}
\put(50,50){\vector(-3,-2){42}}
\put(0,30){$\psi_3$}
\end{picture}
\caption{The three input states $\psi_1$, $\psi_2$, $\psi_3$ 
in the second example. The
optimal measurement is a POVM whose elements are projectors onto the
three states $\psi_1$, $\psi_2$, $\psi_3$.}
\end{figure}
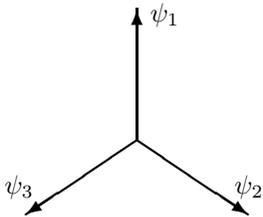

Using the techniques of section \ref{fidelity}, one can show that 
the elements of an optimal POVM are necessarily proportional to the
three projectors $|\psi_1\rangle \langle \psi_1| ,
|\psi_2\rangle 
\langle \psi_2| , |\psi_3\rangle \langle \psi_3|$, see figure 3.
Therefore the optimal POVM whose output entropy is minimum is $\{ 
{2 \over 3} |\psi_1\rangle \langle \psi_1| , {2 \over 3}
|\psi_2\rangle 
\langle \psi_2| , {2 \over 3} |\psi_3\rangle \langle \psi_3|\}$.
In this case $I_{output}^S = \ln 3 > 1$ bits. The other optimal
measurements have larger $I_{output}^S > \ln 3$ bits.
One can also show that there is no measurement on blocks of $L$ input
states whose fidelity is strictly 
equal to the optimum and whose output entropy is less then $L\ln
3$ bits. However if one only requires that the fidelity is arbitrarily
close to the maximum, then in the asymptotic limit ($L\to \infty$)
the output entropy can be made arbitrarily close to $L$ bits, thereby
attaining the bound eq. (\ref{S}). The main difficulty of the proof
will be to construct such a measurement on large blocks
whose output entropy is equal to $L$ bits 
and whose fidelity is arbitrarily close to the optimal
fidelity. 

\section{Plan of the Proof}

The main part of this paper is devoted to proving the bound eq. (\ref{S}).
In section \ref{fidelity} we introduce a large class of fidelities,
and derive some properties of the optimal measurements. In section
\ref{Fid} we show how to generalize these fidelities to measurements
on large blocks of input states. At the end of section \ref{Fid} we
are in a position to state with precision a first version of our main result, 
eq. (\ref{S}).
In section \ref{other} we extend the notion of fidelity and state a
slightly more general version of our result.
In section \ref{comp} we show how to construct a measurement on large
blocks which has little redundancy. In section \ref{optimal} 
we derive an intermediate result
concerning the fidelity of the measurement constructed in section
\ref{comp}. If the states are uniformly distributed in Hilbert
space (ie. the density matrix is proportional to the identity, $\rho =
I/d$) , then this intermediate result already proves our main claim
eq. (\ref{S}). When the states are not uniformly distributed in
Hilbert space, we must first project blocks of states onto the probable
subspace before using the intermediate result of section
\ref{optimal}. This is done in section \ref{prob} and completes
the proof  of eq. (\ref{S}).

\section{Fidelity}\label{fidelity}

Let us consider the general setup described in figure 1.
The states emitted by the quantum source
$|\psi_i\rangle$ belong to a Hilbert space of dimension $d$.
They occur with probability $p_i$.  
Their density matrix is $\rho = \sum_i p_i |\psi_i\rangle \langle \psi_i|$
with $Tr \rho =1$.
The most general measurement on the input states is a
POVM with $M$ element:
 $a_j \geq 0$, $\sum_{j=1}^M a_j = I_d$.

We introduce the fidelity in the following way.
To each outcome $j$ of the measurement we associate a state 
$|\phi_j^{guess}\rangle $
which is our ``guess'' as to what the input  state was.
The correctness of this guess is measured by a function, the fidelity, 
which depends on the input state and the guessed state
$f(|\phi_j^{guess}\rangle ,|\psi_i\rangle )$. 
For instance $f$ could have the form eq. (\ref{1}) or (\ref{3}).
The mean fidelity is then:
\begin{equation}
F = \sum_i p_i \sum_j p(j|i)
f(|\phi_j^{guess}\rangle ,|\psi_i\rangle )\ .
\label{F}\end{equation}
where the probability to obtain outcome $j$ if the state is $ |
\psi_i\rangle  $ is
\begin{equation}
p(j|i) =  \langle \psi_i| a_j | \psi_i\rangle
\end{equation}
An optimal measurement is one which maximizes the mean fidelity $F$.

This is a rather general formulation of the state estimation
problem. 
However the fidelity  
is not the most general one could consider. 
To see this let us consider the optimization of $F$. 
When we make the optimization, we must compare the value of $F$ for
different POVM's,
 {\em however
the guessed states $|\phi_j^{guess}\rangle $ are kept fixed.} That
is the guessing strategy is fixed once and for all, and we try to
optimize the measurement for fixed guessing strategy. The advantage of
formulating the fidelity in this way is technical: it ensures that the
fidelity depends linearly on the POVM elements. 
We shall show in section \ref{other} how to extend our result to more
general fidelities for which the guessed states
$|\phi_j^{guess}\rangle $ are not kept fixed.

We summarize here the main properties of optimal measurements for the
fidelity eq. (\ref{F}), see also \cite{H2}\cite{MP}.

First of all note that we can always take the optimal POVM to consist of one
dimensional projectors $b_j = |b_j\rangle \langle b_j|$ (The $b_j$
are not normalized). Indeed refining a POVM can only increase the
fidelity. This can be seen formally in the following way:
suppose the $a_j$ are an optimal POVM, but not necessarily made out of
one dimensional projectors. Then each $a_j$ can always be decomposed as
$a_j = \sum_k |b_{jk}\rangle \langle b_{jk}|$ since it is a positive
operator.
Inserting this into the expression for $F$ one sees that the $b_{jk}$
(to which we associate the guessed state $\phi_j^{guess}$) are also optimal.

Thus we can optimize $F$ in the class of POVM's whose elements are one
dimensional projectors $|b_j\rangle \langle b_j|$.
These projectors are subject to the unitarity
condition
$\sum_j |b_j\rangle \langle b_j| =I_d$. 
 This can be
implemented by introducing $d^2$ Lagrange multipliers
$\lambda_{\mu\nu}$ which we group into one operator $\hat \lambda$:
\begin{eqnarray}
F & =&  \sum_i p_i \sum_j\langle \psi_i | b_j\rangle \langle b_j |
\psi_i\rangle  f(\psi_i, \phi_j^{guess})
-  Tr [ \hat \lambda (\sum_j| b_j\rangle \langle b_j | - I_d)]\nonumber\\
&=& \sum_j Tr [ (\hat F_j - \hat \lambda) | b_j\rangle \langle b_j | ]
+ 
Tr \hat \lambda\ ,
\label{FFF}\end{eqnarray}
where $\hat F_j = \sum_i p_i | \psi_i\rangle  \langle \psi_i | 
f(\psi_i, \phi_j^{guess})$.
If we vary this with respect to $\langle b_j|$, we obtain the equations 
\begin{equation}
 (\hat F_j - \hat \lambda) | b_j\rangle  =0\ .
\label{flb}
\end{equation}
Inserting this into eq. (\ref{FFF}) shows that $F= Tr \hat \lambda$.

Eq. (\ref{flb}) is the essential equation to find 
optimal measurements explicitly. For instance consider the first example of
section \ref{example}. 
There are two input states $\psi_1$ and $\psi_2$ and two
guessed states $\phi^{guess}_1 = |\psi_1\rangle $
and  $\phi^{guess}_2 = |\psi_2\rangle $. If the input state is 
$|\psi_1\rangle $ and
one guesses $\phi^{guess}_1$, then $f=+1$, whereas if 
the input state is $|\psi_2\rangle $ and
one guesses $\phi^{guess}_1$, then $f=-1$, hence
$\hat F_1 = {1\over 2} (|\psi_1\rangle \langle \psi_1| - 
|\psi_2\rangle \langle \psi_2|)= +\alpha
\beta \sigma_x$.
Similarly $\hat F_2 = {1\over 2} (|\psi_2\rangle \langle \psi_2| - 
|\psi_1\rangle \langle \psi_1|)= -\alpha
\beta \sigma_x$. The task is then to find an operator $\hat \lambda$
such that null eigenvectors of $\hat F_{1,2} - \hat \lambda$
can satisfy the completeness relation. The only possibility is 
$\hat \lambda = \alpha \beta
I$. Therefore the optimal measurement is along the $x$ axis, and
$F_{max} = 2  \alpha \beta$. The second example of section
\ref{example} can be treated along similar lines.

An important 
consequence of eq. (\ref{flb}) is an explicit
expression for the value of $F$ if the measurement is not optimal.
Consider a measurement $a'_j$ which is not optimal, but each
positive operator $a'_j$ is ``close'' to the corresponding operator
$b_j$ of the optimal measurement.
We then decompose the operator $a'_j$ in terms of its components along
$|b_j\rangle $:
$a'_j = X_j |b_j\rangle \langle b_j| + Y_j|b_j\rangle \langle
b_j^\perp| + 
Y^*_j |b_j^\perp\rangle \langle b_j| + 
z_j$ where the state $|b_j^\perp\rangle $ is orthogonal to $|b_j\rangle $ and
the operator $z_j$ obeys $z_j |b_j\rangle  =0$, $\langle b_j| z_j =0$.
Inserting this decomposition into the expression for $F$, we
obtain
\begin{eqnarray}
F(a') &=&  Tr  \hat \lambda + \sum_j 
Tr [ (\hat F_j - \hat \lambda) a'_j]
\nonumber\\
&=& F_{max} + \sum_j Tr [ (\hat F_j - \hat \lambda)  z_j]
\nonumber\\
&\geq&  F_{max} - C \sum_j Tr z_j \ ,\label{boundF}
\end{eqnarray}
where we have used eq. (\ref{flb}) and $C$ is some positive constant
independent of $j$.
This expresses in a simple way how much the fidelity differs from its
maximal value in terms of how much the measurement differs from the
optimal measurement.

\section{Fidelity for measurements on large blocks}\label{Fid}

As discussed above it is necessary to also consider measurements on
large blocks of $L$ input states $| \psi_{i_1}
... \psi_{i_L}\rangle $. The fidelity for measurements on
large blocks is 
\begin{equation}
F_L = \sum_{i1,...,i_L}  p_{i_1}...p_{i_L}
\sum_{j=1}^N \langle  \psi_{i_1} ... \psi_{i_L}| A_j | \psi_{i_1}
... \psi_{i_L}\rangle  {1 \over L} \sum_{k=1}^L f(\psi_{i_k} , 
\phi^{guess}_{j_k}) \ ,
\label{FL}
\end{equation}
where $A_j$ is the measurement on the $L$ input states.
The guessed state is the product $|\Phi_{j}^{guess}\rangle  = 
|\phi_{1_j}^{guess}
... \phi_{L_j}^{guess}\rangle $. The fidelity is taken to be the average of
the fidelities for each state $| \psi_{i_1}\rangle ,...,|\psi_{i_L}\rangle $. 
This ensures
that eq. (\ref{FL}) is just the average of the fidelities
eq. (\ref{F}), as can be seen by rewriting $F_L$ as
\begin{eqnarray}
F_L &=&  {1 \over L} \sum_{k=1}^L \sum_{j=1}^N \sum_{i_k} p_{i_k}
 \langle  \psi_{i_k}| A_j^{(k)} | \psi_{i_k}\rangle  f(\psi_{i_k} ,
 \phi^{guess}_{j_k}) \ ,
\end{eqnarray}
where the 
operators $A_j^{(k)}$ are the operators $A_j$ 
restricted to the space of particle $k$:
\begin{eqnarray}
A_j^{(k)}&=& Tr_{l \neq k} \left( \prod_{l' \neq k}\rho_{l'}\right)
A_j \ .
\label{FLtwo}
\end{eqnarray}

Note that a  possible measurement that maximizes $F_L$ is built out of 
the measurement $\{ a_i \}$ which maximize eq. (\ref{F}):
\begin{equation}
A_j = a_{j_1}\otimes ...\otimes a_{j_L} \ .
\label{optt}
\end{equation}
This measurement has $M^L$ outcomes. And in general $M$ will be larger
than $2^{I_{input}^{VN}}$.

Our main result is that one can always construct optimal measurements with
$2^{I_{input}^{VN}}$ outcomes per input state which also maximize $F$. 
Stated with precision we shall
prove the following result:

Consider a state estimation problem in which the unknown state
$|\psi_i\rangle $ have density matrix $\rho = \sum_i p_i
|\psi_i\rangle  
\langle \psi_i|$
and von Neumann entropy $I_{input}^{VN} = - Tr \rho \ln \rho$.
The quality of the state estimation is measured by a fidelity of the
form eq. (\ref{F}).
Given any $\epsilon > 0$ and $\eta > 0$, then there exists $L_0$
such that for any $L \geq L_0$, and any
$N$ larger than
$2^{L (I_{input}^{VN} +\eta)}$,  
there exists a measurement on sequences of $L$ input states
which has $N$ outcomes and attains a fidelity $F_L \geq F_{max} -
\epsilon$. The Shannon 
entropy of the outputs per input state, $I_{output}^S$, 
can therefore
be made equal or less then
$ I_{input}^{VN} + \eta $. 

It is this result that will be proven in sections \ref{comp} to \ref{prob}.

\section{Other Fidelities}\label{other}

Our main result, as stated with precision at the end of the preceding 
section, applies only to
fidelities of the form eq. (\ref{F}) with fixed guessed states. In this
section we enquire whether it can be generalized to other fidelities?

As  a first generalization, we consider fidelities of the form
eq. (\ref{F}), but for which both the POVM elements $\{ a_j\}$, 
and the guessed
states are undetermined and must be varied to find the optimum
estimation strategy. That is whereas in section \ref{fidelity} the
specification of an estimation strategy consisted only of the POVM
elements $\{ a_j\}$, it now consists of the set $\{ a_j ,
\phi_j^{guess}\}$ which comprises both the POVM elements and the
guessed states. An example of such more general fidelities was
considered in \cite{MP}. The unknown states $|\psi_i\rangle$ where taken 
to be $n$ spin 1/2 particles all polarized along the same direction
$\Omega$ and the fidelity was taken to be the scalar product of one
spin polarized along $\Omega$ with one spin polarized along
the guessed direction $f=|\langle \uparrow_\Omega |
\uparrow_{\Omega_{guess}}\rangle|^2$. 

It is easy to show that our main result eq. (\ref{S}) 
also applies to such more general
fidelities for which both the POVM elements and the guessed states
can be varied. First note that one can always find an optimal 
estimation strategy with only a finite number $M$ of outcomes
\cite{DBE,LPT}. Associated to each outcome is a guessed state
$\phi_j^{guess (OPT)}$, $j=1,...,M$. Let us now consider the subclass
of estimation strategies $\{ a_j ,
\phi_j^{guess(OPT)}\}$ for which the guessed states are fixed to be an 
optimal set and only the POVM elements can vary. 
Note that the optimal fidelity for this subclass is equal 
to the optimal fidelity for the more general estimation strategy since 
the guessed states are taken to be optimal.
Since for this subclass only the POVM elements can vary,
we are in the conditions of section \ref{fidelity} and \ref{Fid}. 
The result
stated at the end of section \ref{Fid} therefore applies.
Hence there
exists a measurement on large blocks whose output entropy is less or
equal to the von Newmann entropy of the input states and whose
fidelity is greater then the optimal fidelity minus $\epsilon$. 
This shows that our main result also holds for these more general
fidelities.

One can however construct even more general fidelities (for instance
by taking the fidelity to be non linear in the POVM elements). For
such more general fidelities it is an open question whether our claim
also applies. One example of such more general fidelities is the
mutual information eq. (\ref{4}). For this particular example our claim 
also holds. This is discussed in the next section.

\section{Relation to the classical capacity of a quantum channel}\label{conc}

In the state estimation problem as presented in figure 1, the
classical source specifies in a completely random manner which quantum
state is emitted. The task of the measurement is to recognize as well as
possible which state was emitted by the quantum source. It is
instructive to compare this to the problem
of classical communication through a quantum
channel\cite{H}\cite{HJSWW}. In this case the classical source chooses 
a controlled subset of all possible sequences 
(called code words) in such a way that they can be recognized  (almost)
perfectly by the receiver. He can then communicate classical
information
reliably through
the quantum channel. 
The relation between the two problems is that in the communication
problem the receiver must recognize the code words, so he is
confronted with a state estimation problem, although it is a
particular one.

For this reason the two problems are related both conceptually and formally.
On the conceptual side,  a corollary of our main result is 
an alternative proof of Holevo's upper
bound on the classical capacity of a quantum  
channel\cite{H} in the case where the quantum channel consists of pure
states. Indeed if the message is to be
transmitted faithfully, Bob must recognize the code words with high
fidelity. We can now view the code words as the states
$|\psi_i\rangle$ that are emitted by the quantum source in figure
1. The von Newmann entropy of the words is less than $n I^{VN} (\rho)$
where $n$ is the number of letters in a word and $I^{VN} (\rho)$ is
the von Newmann entropy of the letters. 
Recall now that
the question answered in this paper is to find, among all
the measurements which  recognize the input words with high
fidelity, those whose output has the minimum entropy. Clearly this
minimum entropy is an upper bound to the capacity of the channel. We
have shown that
it is less or equal to the von 
Neumann entropy of the channel. Thus the quantum channel has a
classical capacity less than $I^{VN} (\rho)$ bits per word, confirming
Holevo's result.

On the formal side, the techniques we have used to construct a
measurement which minimizes the entropy of the outputs are closely
related and inspired by the techniques used to construct a decoding
measurement which maximizes the capacity of the channel \cite{HJSWW}. 
There is
however a very important 
difference with the communication problem. Indeed  in that
case one
can easily build a measurement with a small number of outcomes
(corresponding to a few code words, ie. to a small capacity), and the
task is to try to {\em maximize} the number of outcomes of the measurement
while continuing to recognize the code words faithfully. In this paper
we can
easily build a measurement with a high
fidelity (ie. which is optimal), but with a large redundancy in the
output. The difficulty is to {\em minimize} the number of
outcomes (the redundancy) while keeping the measurement
optimal. Nevertheless the
mathematical technique that
we use in section \ref{comp} to decrease the number of outcomes 
without substantially modifying the measurement is related to the techniques used
 in
\cite{HJSWW}.

\section {Eliminating redundancy}\label{comp}

Our aim in this section is to construct a measurement with less
outcomes than the optimal measurement eq. (\ref{optt}). 
The next two sections will be devoted to
prove that this measurement does not diminish the fidelity.
This  measurement is very similar to the
measurement used in \cite{HJSWW} to decode a classical message sent
through a quantum communication channel. 

We start from the optimal  POVM  acting on one input state and
decomposed into one dimensional projectors $b_i = |b_i\rangle \langle b_i|$.
We express it   in terms of the
normalized operator $\tilde b_i = |\tilde b_i\rangle \langle \tilde
b_i| = b_i / Tr (b_i)$ as  $b_i = \beta_i \tilde b_i$. 
(Throughout the text we shall denote normalized
operators by ${\tilde{ }}\ $). The $\beta_i$ sum to $\sum_i
\beta_i = d$ obtained by taking the trace of the completeness
relation.

We now construct $N$ operators acting on the space of $L$ input states:
\begin{equation}
\tilde B_j= |\tilde B_j\rangle \langle  \tilde B_j| = \tilde b_{j_1}
\otimes \ldots \otimes \tilde b_{j_L}
\label{Bj}
\end{equation}
where each $ \tilde b_{j_k}$ is chosen randomly and independently from the set
$\tilde b_1, \ldots , \tilde b_M$ 
with probabilities $p_1= \beta_1/d ,..., p_M = \beta_M /d$.

The $|\tilde B_j\rangle $ span a subspace $H_B$ of the Hilbert space
of 
the $L$ input
states. In this subspace the operator $B=\sum_j \tilde B_j$ is strictly
positive, hence we can construct the operators
\begin{equation}
C_j = |C_j\rangle \langle C_j| = B^{-1/2} \tilde B_j B^{-1/2}\ .
\label{C}
\end{equation}
The $C_j$ are positive operators,
which sum up to the identity in $H_B$:
$\sum_{j=1}^N C_j = \Pi _{B}$ where $\Pi_B$ is the projector onto
$H_B$. 
The POVM we shall use consists             of the
$C_j$ and  the projector onto the complementary subspace 
$C_0=I_{d^L} - \Pi_{B}$ ($I_{d^L}$ is the identity on the Hilbert
space of the $L$ input states).
 
Our strategy in the next sections 
will be to compute the average fidelity $\overline {F_L }$, 
where the average is taken over possible choices of $B_j$ in
eq. (\ref{Bj}). We shall show that the average of $F_L$ satisfies 
our main result stated at the end of section \ref{Fid}.
Therefore 
there necessarily are some choices of $B_j$ which also satisfy our
main result.

But first we 
 derive some important
properties of the $C_j$. We shall obtain mean properties, where the
mean is the average over choices of $B_j$ in eq. (\ref{Bj}).
\begin{itemize}
\item
The mean of $\tilde B_j$ is $\overline{ \tilde B_j} = I_{d^L} / d^L$. 
\item
The mean of $B$ is:
\begin{eqnarray}
\overline{ B }
&=& \sum_{j=1}^N  \overline{\tilde B_j } =
 {N \over d^L} I_{d^L} \ .
\end{eqnarray} 
This motivates our writing 
\begin{equation}
B =  {N \over d^L}\left ( I_{d^L} +
\Delta \right)
\end{equation} and subsequently making expansions in $\Delta$.
\item
The dimension of $H_B$ is
\begin{eqnarray}
dim_{H_B} &=& \sum_j Tr C_j =\sum_j Tr B^{-1} \tilde B_j 
=  {d^L \over N} \sum_j Tr {1 \over  I_{d^L} +
\Delta } \tilde B_j \nonumber\\
&\geq&  {d^L \over N}
 \sum_j Tr (I_{d^L} - \Delta   ) \tilde B_j \ .
\end{eqnarray}
Furthermore
\begin{eqnarray}
Tr  \Delta   \tilde B_j &=& 
 Tr ({d^L \over N} B - I_{d^L} ) \tilde
B_j \nonumber\\
&=& Tr \left[ {d^L \over N} ( \tilde B_j  +\sum_{k \neq j} 
 \tilde B_k  \tilde B_j ) - \tilde B_j \right]\end{eqnarray}
where we have used the fact that $ \tilde B_j^2 = \tilde  B_j$.
We now take the average of this expression. 
 Using the fact that for $k \neq j$, $ \tilde  B_k$
 and $ \tilde B_j$ are independent, the average of  $ \tilde B_k
 \tilde 
 B_j (k \neq j)$ is
the product of the averages$ \overline{  \tilde B_k  \tilde B_j } =
 \overline{
 \tilde
 B_j } \ \overline{
 \tilde B_k } =I_{d^L} / d^{2L}
 $. And hence $\overline {\sum_{k \neq j} 
 \tilde B_k  \tilde B_j } = (N-1) I_{d^L} / d^{2L}$.
Putting all together, we find $\overline{ {Tr  \Delta   \tilde  B_j}}
= {d^L -1 \over N} $ and 
\begin{equation}
d^L \geq \overline{ {dim H_B} }
\geq d^L (1 - {d^L -1 \over
N}) \ .
\label{dim}\end{equation}
This shows that if $N$ is slightly larger than the dimension of the
Hilbert space $d^L$, 
then the $C_j$ ($j \neq 0$) fill the Hilbert space.

\item
Finally we need to know how much the $C_j$ differ from the $\tilde B_j$. We
write $|C_j \rangle  = \alpha_j |\tilde B_j\rangle  +
|B_j^\perp\rangle $ 
and compute $\alpha^2_j$: 
\begin{eqnarray}
 \alpha^2_j &=&  Tr C_j\tilde B_j \nonumber\\
&=&  Tr\tilde B_j B^{-1/2}\tilde B_j B^{-1/2} \nonumber\\
&=& \left( Tr \tilde B_j B^{-1/2} \right)^2 \nonumber\\
&\geq&{d^L \over N} \left( 1 -  {1\over 2}Tr\tilde B_j \Delta
\right)^2
\ .
\end{eqnarray}
Hence
\begin{eqnarray}
\overline{\alpha^2_j }
&\geq& {d^L \over N} ( 1 - \overline{ Tr\tilde B_j \Delta })\nonumber\\
&=&  {d^L \over N} (1 - {d^L -1 \over N})\ .
\end{eqnarray}
This is then used to compute the average of $\langle
B_j^\perp|B_j^\perp
\rangle $:
\begin{equation}
\overline{\langle B_j^\perp|B_j^\perp\rangle  }= \overline{ Tr C_j } - 
\overline{ Tr C_j B_j } 
\leq  {d^L \over N} {d^L -1 \over N}\ ,
\label{betaj}\end{equation}
which shows that the $C_j$ are arbitrarily close to the $\tilde B_j$
when $N > d^L$.
\end{itemize}

\section{An intermediate result}\label{optimal}

In this section we shall prove the following intermediate result:

Suppose that the input states  $|\psi_i\rangle $
belong to a Hilbert space of dimension $d$
and have a density matrix $\rho = \sum_i p_i |\psi_i\rangle
\langle 
\psi_i| $. 
Denote by $\rho_{max}$
the largest eigenvalue of $\rho$. Consider measurements on blocks
of $L$ input states.
Give  yourself any positive number
$\eta >0$. 
Let  
$N$ be any integer larger than
$2^{L (2\ln d + \ln \rho_{max} +\eta)}$. Then
there exist measurements  
with  $N$ outcomes with a
fidelity
$F_L \geq F_{max} - R 2^{- L \eta}$ where $R$ is a positive constant.

In the next section we shall combine this intermediate result with the
concept of probable subspace of a long sequence of states to prove our 
claim in full generality.

To prove this intermediate result, we proceed as follows:

Let $\{ b_j=|bj\rangle \langle b_j| \}$ be a POVM that maximizes the 
fidelity $F$
eq. (\ref{F}). Using the algorithm of eq. (\ref{Bj}) to (\ref{C}) we
construct a measurement $C_j$ , $j=0,...,N$ acting on the space of $L$ copies
of the input states.

Let us consider the fidelity for the measurement $C_j$:
\begin{eqnarray}
F_L &=&       \sum_{j=0}^N  {1 \over L} \sum_{k=1}^L \sum_{i_k}  p_{i_k}
\langle  \psi_{i_k}| C_j^{(k)} | \psi_{i_k}\rangle  
f(\psi_{i_k} , \phi^{guess}_{j_k})
\label{Smult2}
\end{eqnarray}
where the $C_j^{(k)} = Tr_{l \neq k} \left( \prod_{l' \neq
k}\rho_{l'}\right) C_j$  are defined as in
eq. (\ref{FLtwo}).  

We can decompose $C_j^{(k)}$ (for $j \neq
0$) according to
its components along $|\tilde b_{jk}\rangle $:
$C_J^{(k)}= X_{jk} |\tilde b_{jk}\rangle \langle \tilde b_{jk}|
+ Y_{jk} | \tilde b_{jk}\rangle \langle  b_{jk}^\perp| + 
Y_{jk}^* | b_{jk}^\perp\rangle \langle \tilde b_{jk}|
+ z_{jk}$
where $z_{jk} |\tilde b_{jk}\rangle  = 0$, $\langle \tilde b_{jk}|z_{jk} =0$.
Inserting this expression in  eq. (\ref{Smult2}), and using
eq. (\ref{boundF}),  yields
\begin{eqnarray}
F_L
&\geq& {1 \over L} \sum_{k=1}^L \left (
 F_{max}  - C  \sum_{j=1}^N Tr  z_{jk}
- C Tr C_0^{(k)}\right ) 
\end{eqnarray}
where the last term comes from the $C_0 = I_{d^L} - \Pi_B$ outcome.

It remains to calculate $Tr C_0^{(k)}$ and $Tr z_{jk}$. 
We start with the former
\begin{eqnarray}
C_0^{(k)} &=&  Tr_{l \neq k} (\prod_{l' \neq k} \rho_{l'}) (I_{d^L} -
\Pi_B)\nonumber\\
&\leq & (\rho_{max})^{L-1} Tr  (I_{d^L} -
\Pi_B) =  (\rho_{max})^{L-1} (d^L - dim \ H_B)\label{C0}\end{eqnarray}
where $\rho_{max}$ is the largest eigenvalue of $\rho$.

To estimate  $Tr z_{jk}$
we recall the decomposition of 
$|C_j\rangle  = \alpha_j |\tilde B_j\rangle  + |B_j^\perp\rangle $.
We can further decompose $|B_j^\perp\rangle $ according to whether when
restricted to the space of the $k$'th particle, it is equal to 
$|b_{j_k}\rangle $ or
not:
$|B_j^\perp\rangle $ = $|\tilde b_{j_k}\rangle |\phi\rangle  + 
|\tilde b_{j_k}^\perp\rangle |\chi\rangle $.
Inserting this into the trace which yields $C_j^{(k)}$, we obtain
\begin{eqnarray}
C_j^{(k)}& =& Tr_{l \neq k}(\prod_{l' \neq k} \rho_{l'})
\left (\alpha_j| \tilde B_j\rangle 
+ | \tilde b_{j_k}\rangle |\phi\rangle  + | \tilde
b_{j_k}^\perp\rangle 
|\chi\rangle  \right)
\left ( \alpha_j^*
\langle \tilde B_j | +...
..\right)\nonumber\\
&=& | \tilde b_{j_k}\rangle \langle  \tilde b_{j_k}| X_{jk} 
+  | \tilde b_{j_k}\rangle \langle  \tilde b_{j_k}^\perp| Y_{jk} + | \tilde
b_{j_k}^\perp\rangle \langle  \tilde b_{j_k}|Y_{jk}^*
+ | \tilde b_{j_k}^\perp\rangle \langle  \tilde b_{j_k}^\perp| Z_{jk}
\ .
\end{eqnarray}
The coefficients $X_{jk}$, $Y_{jk}$, $Z_{jk}$ are easily calculated.
The one of interest is $Z_{jk} = Tr z_{jk}$:
\begin{eqnarray}
Z_{jk} &=& Tr \prod_{l' \neq k} \rho_{l'}
|\chi\rangle \langle \chi|
\nonumber\\
&\leq & (\rho_{max})^{(L-1)} 
\langle \chi|\chi\rangle \nonumber\\
&\leq & (\rho_{max})^{(L-1)} \langle B_j^\perp|B_j^\perp\rangle 
\ .
\end{eqnarray}

Inserting these bounds  into the expression for $F_L$ we obtain
\begin{eqnarray}
F_L
&\geq&  {1 \over L} \sum_{k=1}^L \left ( F_{max}  - {C (\rho_{max})^{L-1}}
 \langle  B_j^\perp |B_j^\perp\rangle  -  {C (\rho_{max})^{L-1}} 
(d^L - dim H_B) \right)\ .
\label{bound}
\end{eqnarray}
We now take the average of this expression over all possible choices
of $b_{jk}$ operators in eq. (\ref{Bj}). 
Inserting eq. (\ref{dim}) and (\ref{betaj}) yields
\begin{equation}
\overline{F_L} \geq F_{max}
-2 C (\rho_{max} )^{L-1} d^L{d^L - 1 \over N} \ .
\end{equation}
 Therefore  if $N\geq 2^{L(2\ln  d + \ln \rho_{max} + \eta)}$, then
$\overline{F_L}  \geq F_{max} - R 2^{-L \eta}$ where $R=2 C /\rho_{max}$. 
This proves the intermediate result.

Note that 
if the input states are uniformly distributed in Hilbert space, ie.
$\rho = I/d$,  then this intermediate result 
directly implies our main claim. Indeed when $\rho = I/d$,
$\rho_{max} = 1/d$, then $\overline{F_L} \geq F_{max} - R 2^{-L\eta}$ if
$N\geq 2^{L(\ln  d + \eta)}
= 2^{L(I_{input}^{VN} + \eta)}$. 
When the input states are not uniformly distributed in Hilbert space,
we must use the notion of probable Hilbert space of a long sequence
to prove our main result. This is done in the next section.

\section{Measurements on probable subspaces}\label{prob}

We now combine the result of the previous section with the notion of
probable subspace of large blocks of states.

We first recall 
the properties of the probable subspace\cite{S}\cite{JS}.
Consider a long sequence of $L'$ 
input states $|\psi_{i_1}...\psi_{i_{L'}}
\rangle $. The density matrix of these
states is $\rho= \prod_{k=1}^{L'}\rho_k$. 
The projector $\Pi$ onto the probable subspace has
the properties that given $\epsilon'>0$, $\eta' >0$, and for $L'$
sufficiently large,

\begin{enumerate}
\item $Tr \Pi \rho \geq 1 - \epsilon'$, ie. the probability to be in
the probable subspace is arbitrarily close to $1$.

\item $\Pi$ and $\rho$ commute, ie. the eigenvectors of $\rho$
are either eigenvectors of $\Pi$ or of $1-\Pi$.
And furthermore the eigenvectors which are common to $\Pi$ and
$\rho$ have eigenvalues comprised between
$2^{L'(-H - \eta')}\leq  (\rho_{L'})_i \leq 2^{L'(-H + \eta')}$

\item From these two properties it follows that the dimension of the
probable Hilbert space is bounded by
$( 1 - \epsilon' ) 2^{L'(H - \eta')}\leq Tr \Pi \leq  2^{L'(H +
\eta')}$
\end{enumerate}

Let us now show that measurements restricted  to the probable subspace
are arbitrarily close to optimal.
Suppose that $A_j$ is a measurement that
optimizes the state determination problem eq. (\ref{FL}) for sequences
of $L' $ input states (for instance the measurement eq. (\ref{optt}). 
Consider the POVM consisting of
the operators
$A'_j = \Pi A_j \Pi$ (to which we associate the unmodified 
guessed states $\phi_{j_k}^{guess}$) 
and the operator
$I-\Pi$ (to which we associate the minimal value of the fidelity
$f_{min}$). 
The fidelity for this measurement is

\begin{eqnarray}
F_{L'} &=&
\sum_{i_1 ... i_{L'}} p_{i_1}... p_{i_{L'}} 
\sum_{j=1}^N \langle  \psi_{i_1} ... \psi_{i_{L'}}| 
\Pi A_j \Pi  | \psi_{i_1}
... \psi_{i_{L'}}\rangle  {1 \over L'} 
\sum_{k=1}^{L'} f(\psi_{i_k} , \phi_{j_k})
\nonumber\\
& &+ \sum_{i_1 ... i_{L'}} p_{i_1}... p_{i_{L'}} 
\langle  \psi_{i_1} ... \psi_{i_{L'}}|   1- \Pi 
 | \psi_{i_1}
... \psi_{i_{L'}}\rangle
f_{min}
\nonumber\\
&\geq& F_{max} \nonumber\\ & &
- 
\sum_{i_1 ... i_{L'}} p_{i_1}... p_{i_{L'}} 
\sum_{j=1}^N \langle  \psi_{i_1} ... \psi_{i_{L'}}| 
A_j - \Pi A_j \Pi  | \psi_{i_1}
... \psi_{i_{L'}}\rangle
{1 \over{L'} } \sum_{k=1}^ {L'}f(\psi_{i_k} , \phi_{jk})
\nonumber\\
& &
+  f_{min} Tr \rho (1 - \Pi) \ . 
\end{eqnarray}
We 
bound the second term by
\begin{eqnarray}
&& | \sum_{i_1 ... i_{L'}} p_{i_1}... p_{i_{L'}} 
\sum_{j=1}^N \langle \psi_{i_1} ... \psi_{i_{L'}}| 
 A_j - \Pi A_j \Pi  | \psi_{i_1}
... \psi_{i_{L'}}\rangle 
  {1 \over{L'} } \sum_{k=1}^{L'} f(\psi_{i_k} , \phi_{jk}) | 
\nonumber\\
&\leq&f_{max} \sum_{j=1}^N | 
\sum_{i_1 ... i_{L'}} p_{i_1}... p_{i_{L'}} 
\langle  \psi_{i_1} ... \psi_{i_{L'}}|   ( A_j - \Pi A_j \Pi) 
| \psi_{i_1}
... \psi_{i_{L'}}\rangle 
|
\nonumber\\
&=&f_{max} \sum_{j=1}^N | Tr [ \rho (A_j - \Pi A_j \Pi) ]|  \nonumber\\
&=& f_{max} Tr[ ( \rho - \Pi \rho \Pi) \sum_{j=1}^N A_j] 
=  f_{max} Tr  \rho (I - \Pi
)\nonumber\\
&\leq& \epsilon ' f_{max}
\end{eqnarray}
where $f_{max}$ is the maximum value of the fidelity and
we have used the fact that $\rho - \Pi \rho \Pi$ is a positive operator,
and therefore that $ Tr  [ \rho (A_j - \Pi A_j \Pi) ] \geq 0$ which
allows us to remove the absolute value sign and put the sum over $j$
inside  the trace.

Putting everything together we have
\begin{equation}
F_{L'} \geq F_{max} - \epsilon' (f_{max} - f_{min} )\ .
\label{fmax}\end{equation}
This shows that the restriction of the measurement to the
probable Hilbert space diminishes the fidelity by an arbitrarily
small amount $\epsilon'(f_{max} - f_{min} )$.

We can now build a measurement
which satisfies our main result as stated at the end of section \ref{Fid}.
We decompose the input
states into blocks of $L'$ states. On each of these blocks we first carry
out a partial measurement $\Pi$ and $I - \Pi$ to know whether it is in
the probable subspace or not. If the result is $I- \Pi$ the sequence
is discarded. The sequences which pass the test are kept. 

We now take the sequences which have passed the test 
as the  input states in the
intermediate result. These sequences belong to a Hilbert
space of dimension $dim\ 
H_{probable} \leq 2^{L' (I_{input}^{VN} + \eta')}$ and the largest
eigenvalue of their density matrix is 
$\rho_{max} \leq 2^{L' (-I_{input}^{VN} + \eta')}$.
To apply the intermediate result, 
we take an integer $L$ and an $\eta>0$. Then
 there exists 
a measurement on blocks of $L$ sequences which has 
a number of possible outcomes
equal to any integer $N$ larger than
$2^{L ( L' (I_{input}^{VN} + 3 \eta') + \eta)}
= 2^{L L' ( I_{input}^{VN} +  3 \eta' + \eta /  L')}$
 and which has a fidelity larger than
$F_{LL'} \geq F_{max} -  \epsilon' (f_{max} - f_{min})
- R 2^{- L \eta}$ where $R$ is a positive constant.

Let us calculate the entropy $I_{outputs}^S$
of the outputs of this measurement.
We need less than $I_{\epsilon'} = -\epsilon' \ln \epsilon'
- (1- \epsilon') \ln  (1- \epsilon')$ bits to describe whether or not
the input state passes the first test of belonging to
the probable Hilbert space or not.
If it does then we need less than $\ln N$ 
bits to encode the  output of the measurement on the $L$ blocks of 
probable sequences.
Therefore the total number of bits we need to describe the outcome of
this measurement on $L L' $ elementary input states is
$I_{output}^S \leq  \ln N + L I_{\epsilon'}$. 
Replacing $N$ by its bound, we have
$I_{output}^S \leq  L L'
( I_{input}^{VN} + ( 3 \eta' + \eta /  L' + I_{\epsilon'}/ L'   )$.
Since $\epsilon'$, $\eta'$ and $\eta$ can be chosen arbitrarily
small, and $L'$ arbitrarily large, our claim is proven.

\section{Conclusion}

In this paper we have obtained a quantitative estimate of how much
information can be obtained by a quantum measurement. We considered
optimal measurements, that is measurements which maximize a fidelity
function. We then enlarged the set of optimal measurements in two 
ways. First we considered optimal measurements that act collectively
on large blocks of input states rather than measurements restricted to
act on each state separately. Secondly we did not require the fidelity
of the  measurements to be exactly equal to the optimal fidelity, but
only that it be arbitrarily close to the optimal fidelity. In this
context we showed that  whatever property of a quantum system one wants to
learn about, one can learn at most one bit of information about
every qubit of quantum information carried by the unknown quantum
system. 
That
is, the Shannon entropy of the outcomes of optimal measurements can
always be made equal or less than the von Newmann entropy of the  unknown
quantum states.

\bigskip

\noindent
{\bf Acknowledgments :} 
S.M. would like to thank Utrecht University
where most of this work was carried out. He is a ``chercheur
qualifi\'e'' of the Belgian National Research Fund.


\begin{thebibliography}{999}

\bibitem{Hel} C. W. Helstrom, {\it Quantum Detection and Estimation
Theory}, New York, Academic Press, 1976

\bibitem{H2} A. S. Holevo, {\it Probabilistic and Statistical Aspects
of Quantum Theory}, North Holland, Amsterdam, 1982

\bibitem{MP} S. Massar and S. Popescu, Phys. Rev. Lett. {\bf 74}
(1995) 1259

\bibitem{DBE} R. Derka, V. Buzek, A. K. Ekert, Phys. Rev. Lett. {\bf
80} (1998) 1571

\bibitem{LPT} J. I. Latorre, P. Pascual, R. Tarrach,
Phys. Rev. Lett. {\bf 81} (1998) 1351

\bibitem{LPT2} G. Vidal, J. I. Latorre, P. Pascual, R. Tarrach,
quant-ph/9812068


\bibitem{Sh} C. E. Shannon, Bell. Syst. Tech. J. {\bf 27} (1948) 379

\bibitem{S} B. Schumacher, Phys. Rev. A {\bf 51} (1995) 2738

\bibitem{JS} R. Jozsa and B. Schumacher, J. Mod. Opt. {\bf 41} (1994)
2343


\bibitem{H} A. S. Holevo, Probl. Peredachi Inf. {\bf 9} (1973) 3
[Probl. Inf. Transm. (USSR) {\bf 9} (1973) 177]

\bibitem{HJSWW} P. Hausladen, R. Jozsa, B. Schumacher,
M. Westmoreland, W. K. Wooters, Phys. Rev. A {\bf 54} (1996) 1869


\end{thebibliography}
\end{document}